\documentclass[a4paper]{article}

\providecommand{\preprintno}[1]{}

\def\Version/{1.6}

{\def\$#1: #2 ${#2}}

\def\hepawk/{\.{hepawk}}
\def\f77/{\.{FORTRAN-77}}
\def\hepevt/{\.{/hepevt/}}
\def\KRONOS/{\.{KRONOS}}
\def\KROWIG/{\.{KROWIG}}

\newenvironment{example}[2]{
    \def\examplecaption{#1}
    \def\examplelabel{#2}
    \begin{figure}[htb]
    \large\normalsize
    \begin{tabbing}
    \tab\tab\tab\tab\tab\tab\tab\tab\tab\tab\kill
  }{
    \end{tabbing}
    \caption{\examplecaption}
    \label{\examplelabel}
    \end{figure}
  }

\def\.#1{{\tt#1}}			
\def\,#1{{\rm\it#1\/}}
\def\C{\`\it}                           
\def\tab{\quad\=}
\def\LB{\>\.{\{}\+\+}                   
\def\RB{\<\.{\}}\-\-}                   

\def\Item#1#2{\item[\.{#1}]\hfill#2\goodbreak}

\thicklines

\begin{document}

\title{\hepawk/, Version \Version/ \\
      A Language for Scanning High Energy Physics Events}

\author{%
  Thorsten Ohl%
    \thanks{e-mail: {\tt ohl@crunch.ikp.physik.th-darmstadt.de}}\\
  \hfil\\
    Technical University of Darmstadt\\
    Schlo\ss{}gartenstr. 9\\
    D-64289 Darmstadt\\
    Germany}

\preprintno{IKDA 95/15\\hep-ex/9504007}
\date{April 1995}

\maketitle
\begin{abstract}
  This manual describes version \Version/ of the
  programming language \hepawk/, designed for convenient
  scanning of data structures arising in the simulation of
  high energy physics events.  The
  interpreter for this language has been implemented in \f77/,
  therefore \hepawk/ runs on any machine with a \f77/ compiler.
\end{abstract}


\section*{Program Summary}

\begin{tabular}{lp{8cm}}
Title of the program:   & \hepawk/ \hfil\\
\\
Computer/Operating System: & Any with a \f77/ environment\\
\\
Programming language:   & \f77/\\
\\
Memory required:        & 400k words (including
                          histogramming package)\\
\\
Number of bits in a word: & 32\\
\\
Number of program lines: & $\approx$ 11000\\
\\
Other programs used:    & \begin{raggedright}
                          HBOOK\cite{BL87}, histogramming
                          package from the CERN library.
                          \end{raggedright}\\
\\
Keywords:               & Language translation, Monte Carlo event
                          generation,  High energy physics.\\
\\
Nature of physical problem:
                        & Developing a Monte Carlo event generator for
                          high energy physics requires extensive
                          testing by analyzing its output.\\
\\
Method of solution:     & By implementing an interpreter for a special
                          purpose language dedicated to scanning the
                          output of Monte Carlo event generators,
                          recompilation of the \f77/ program can be
                          avoided.  Since the language has special
                          constructs for scanning high energy physics
                          events, the desired analysis can be
                          specified very concisely.  The
                          interpreter is implemented by using a parser
                          generator and is therefore easy to maintain
                          and very reliable.\\
\\
Typical running time:   & Depending on the application.
                          Comparable to the equivalent straight
                          \f77/ version.
\end{tabular}

\newpage

\section*{Long Write Up}


\section{Introduction}
\label{sec:intro}

Developing a Monte Carlo event generator for high energy physics
involves a substantial amount of testing for different observables and
kinematical regions.  Traditionally this is done by writing a \f77/
subroutine analyzing the event generator's output, which will be
modified, recompiled and relinked for each set of cuts.  To reduce the
number of necessary recompilations, one usually defines some canonical
variables to parameterize the cuts and reads their values from a file.

On the other hand,
the latter approach is not very flexible.  We therefore propose to
generalize the standard approach by using a complete programming
language for specifying the cuts and filling the histograms.
In this case, the
interpreter for the programming language can be linked once and for
all with the Monte Carlo program under consideration.
Furthermore this
programming language allows a very concise specification of the
analyzer, because operators and builtin functions for common
geometrical calculations are provided.

At first sight it might seem to be too much effort to create a new
programming language for a specialized task,
but a closer inspection shows that this is not
the case.  The theory of translation of programming languages is well
established \cite{AU77} and the implementation of
simple programming languages has become a straightforward exercise
(e.g.~\cite{KP84}), provided the available tools are used.

Furthermore,
the emergence of standards \cite{hepevt} for passing event information
between different Monte Carlos allows to create reusable interfaces.
Therefore the interpreter (or compiler) for an event scanning language
can be used unchanged for all Monte Carlos adhering to this standard.
An added advantage of this approach is the  decoupling of the
Monte Carlo generator from the analyzing routine.  No internal
data structures of the generator will be available to the analyzer,
thus the generator is tested as it will appear to the
application program.

This paper describes the implementation of \hepawk/, a language for
scanning high energy physics events.   \hepawk/ was developed as a
back-end for the \KRONOS/ Monte Carlo for radiative corrections at
HERA \cite{ADM+92a} and has been used extensively during the
construction of this generator.

The outline of this write up is as follows:
after an introduction to \hepawk/ in section \ref{sec:design},
we describe
in section \ref{sec:f77} the \f77/ interface.  The implementation
is described in detail in section \ref{sec:implementation} and
section \ref{sec:future} outlines possible further developments.
A reference manual and a test run are presented in the appendix.
This manual is an updated version of \cite{Ohl92a}.

The name \hepawk/ stands for \,{High Energy Physics AWK}.  The latter
is the standard UNIX text manipulation and report generation language
\cite{AKW88}, from which \hepawk/ inherited some of its syntax.


\section{Design}
\label{sec:design}

In this introduction we shall describe the design of \hepawk/ and
shall informally introduce most
of the features of \hepawk/ version \Version/.  We proceed by discussing
an example of a complete \hepawk/ script, introducing concepts in the
order of appearance.  For further details and more complete
descriptions, the reader is referred to the appropriate section of the
reference manual in appendix \ref{sec:ref-man}.

As already mentioned, \hepawk/ has been designed to meet two distinct
purposes: allow concise specifications of analyzers for Monte Carlo
generated events and avoid recompilation and relinking
of large programs.

The latter purpose is of course purely technical.  However, linking
a Monte Carlo event generator with a huge library like the CERN
library takes a considerable amount of time on most computer
systems.  Avoiding this step allows to test a Monte Carlo for
a broader scope of observables.

To meet the former purpose,
\hepawk/ version \Version/ provides the following features:
\begin{itemize}
  \item data structures: \,{vector} and \,{particle}.
  \item \,{vector} operators: \.+, \.-, \.*, etc.
  \item control structures: \.{for}, \.{if} -- \.{else}, \.{while}.
  \item kinematics: \.{angle ()}, \.{rap ()} (rapidity), \.{/|}
    (Lorentz transform), etc.
  \item simple access to histogramming: \.{book1 ()}, \.{fill ()}, etc.
\end{itemize}
The availability of these features should pay off for the additional
effort of having to learn yet another language.  Especially the
richer set of control structures (compared to \f77/) and the
vector expressions have proven to be very convenient.

It should be noted at this point that most of \hepawk/'s features
could easily be implemented in a \.{C++} Monte Carlo library\cite{BL91},
where data structures and overloaded operators are readily
available.

However, all major Monte Carlo event generators to date are written
in \f77/ and interfacing \f77/ to other languages like \.{C++} is
unfortunately machine dependent.  Furthermore, \.{C++} is only
available for a small subset of the computers used in high energy
physics.  In contrast,
\hepawk/ offers a highly portable implementation
of these features.
The same considerations
apply to \.{FORTRAN-90}, although in this case linkage to
\f77/ is guaranteed by the standard.


\subsection{Features}
\label{sec:features}

\hepawk/ scripts are in free format, with two exceptions: everything
from column 73 to the end of the line will be ignored and the
end of the line terminates comments.  The former restriction is
necessary to avoid errors when moving \hepawk/ scripts to computers
which assume fixed size records in their input stream.

In principle, \hepawk/ does {\em not\/} restrict the complexity
of the script.  However, very complex scripts with deeply
nested control structures might exceed the size of \hepawk/'s
run time stack.  In this case, \hepawk/ can easily be recompiled
with a larger stack.   Very long scripts with many variables might
need a larger size for the memory pool managed by \hepawk/, this can
also be achieved by a recompilation.

Each \hepawk/ script is a sequence of conditions and actions.  An
action will be executed for each event for which the condition
evaluates to true (by convention, an empty condition is always true).

\hepawk/ has the builtin datatypes \,{scalar}, \,{vector}, and
\,{particle}.  Vectors are usual Lorentzian four-vectors and the
usual arithmetical operators are defined for them.  Thus the inner
product of two vectors \.{\$p} and \.{\$q} can be assigned to the
scalar \.{pq} by
{\small\begin{verbatim}
   pq = $p * $q;
\end{verbatim}}
The interpretation of the operators \.+, \.-, and \.* is uniquely
determined by their operands.  This feature allows to express
kinematical calculations very concisely.

The output of the event generator is accessed by looping over
subsets of particles in the standard common block.  For example
{\small\begin{verbatim}
   for (@p in ELECTRONS)
     {
        # a group of statements
        # operating on the electron @p
     }
\end{verbatim}}
executes the statements enclosed by the braces for each electron
stored by the generator in the standard common block.

The annotated example in section \ref{sec:example}
is taken from a study of
multi photon events at HERA, using the \KRONOS/ Monte Carlo
\cite{ADM+92a}.  The script addresses two questions: which
multiplicity for observed bremsstrahlungs photons is to be expected
at HERA and what is the energy spectrum of photons in multi photon
events?

To answer these questions, it is of course important to take
into account the detector geometry.  This is accomplished by
specifying angular ranges for the main detector and the two
luminosity monitors.  These can be implemented easily by
conditionals of the type
{\small\begin{verbatim}
   if (theta_min <= angle ($photon, $z_axis) <= theta_max)
     {
        # process the event satisfying
        # the above angular cut
     }
\end{verbatim}}

The multiplicity obviously depends on the minimal energy of
observed photons.  In order to collect the multiplicity for
several cuts, we have to loop over these cuts, using a
\.{while} iteration:
{\small\begin{verbatim}
   E_cut = E_cut_min;
   while (E_cut <= E_cut_max)
     {
       if ($photon:E >= E_cut)
         {
            # process the event satisfying
            # the above energy cut
         }
       E_cut = E_cut + E_cut_step;
     }
\end{verbatim}}


\subsection{An annotated example}
\label{sec:example}

\subsubsection{The \.{BEGIN} action}
\label{sec:example:BEGIN}

The \.{BEGIN} action is executed before the first event is processed.
Typically it defines some variables and allocates histograms.
There can be several \.{BEGIN} actions in a \hepawk/ script, they will
be concatenated.

This {\tt BEGIN} action starts by printing some information on
the Monte Carlo run which will be processed.  The builtin function
\.{printf} is similar to the function with the same name from the
standard \.C library.  See section \ref{sec:output} for details.
{\small\begin{verbatim}
# multiphoton.hepawk

BEGIN
  {
    printf ("\nKRONOS, %s:\n", REV);
    printf ("Run: %d,  Date: %s\n\n", RUN, DATE);

\end{verbatim}}

The builtin variables \.{REV}, \.{RUN}, and \.{DATE} are
preset to the version and revison date of the Monte Carlo,  the
number of the run, and the date of the run respectively.  This
information is taken from \hepevt/.  See section
\ref{sec:autovar} for more automatic variables.

Next we introduce variables as symbolic names for kinematical cuts,
detector geometry, and histogramming parameters.
{\small\begin{verbatim}
    x_min = 0.001; x_max = 0.1; Q2_min = 100;   # kinematical cuts

    E_min = 0.5; E_max = 40;                    # histogramming range
    n_max = 6; E_cut_max = 3; E_cut_step = 0.1; # multiplicity cuts
    E_cut_bins = E_cut_max / E_cut_step;

    th_min_lumi = 0; th_max_lumi = 0.5e-3;      # luminosity monitors
    th_max_main = PI - 0.100;                   # main detector (fwd)
    th_min_main = 0.150;                        # (bwd)

    th_sep_jet = 0.300;                         # separation from jet
    th_sep_em  = 0.150;                         # sep. from electron

\end{verbatim}}

Finally we allocate counters and histograms for the various event
types and observables we want to study.  See section
\ref{sec:histogramming} for more details on the interface to the
histogramming package.
{\small\begin{verbatim}
    incut_0 = 0;                             # counting 0 photon evts

    incut_1 = 0;                             # counting 1 photon evts
    h_E = book1 (0, "dsigma/dE", E_max, 0, E_max);

    incut_2 = 0;                             # counting 2 photon evts
    h_E1 = book1 (0, "dsigma/dE1", E_max, 0, E_max);
    h_E2 = book1 (0, "dsigma/dE2", E_max, 0, E_max);

                                             # Multiplicities
    h_m = book2 (0, "Multiplicity", n_max, - 0.5, n_max - 0.5,
                                    E_cut_bins, 0, E_cut_max);

    h_E1_E2 = book2 (0, "dsigma/dE1dE2",     # Lego plot
                     E_max, 0, E_max, E_max, 0, E_max);
  }
\end{verbatim}}

\subsubsection{Kinematics}
\label{sec:example:kinematics}

The calculation of the basic kinematical variables should be
almost self explaining.
This action will be executed for {\em all\/} events, because
no selection condition is given.

The basic data types in \hepawk/ version \Version/ are \,{real},
\,{vector}, and \,{particle}.  The lexical convention is to use
identifiers starting with \.{\$} for vectors and identifiers
starting with \.@ for particles.  Components of these data structures
are addressed by trailing selections, such as \.{:p} for the four
momentum of a particle.  Therefore the first line in the following
action calculates the Mandelstam variable $s$ from the four momenta
of the two beam particles \.{@B1} and \.{@B2}.  See section
\ref{sec:data-types} for more details on the available data types.
{\small\begin{verbatim}

  {
    S = (@B1:p + @B2:p)^2; # Center of mass energy

\end{verbatim}}

Next we loop over all outgoing electrons and store the momentum
in \.{\$el}.
{\small\begin{verbatim}
    for (@p in ELECTRONS)  # Collect the outgoing electron.
      $el = @p:p;
\end{verbatim}}

This method of obtaining the outgoing electron
momentum is actually quite naive and uses
the fact that the Monte Carlo {\tt KRONOS} will generate one and
only one electron.  A more sophisticated example would have
to use some physical criteria
for distinguishing the scattered electron from
electrons created in hadronic decays.
An example would be to select the electron which is furthest away
from any hadronic jet.  Similar code selecting the hardest photon
can be found in section \ref{sec:example:processing}.

In the next step the electron momentum is used to calculate the
(electronic) momentum transfer and the usual Bjorken variables
$x,y$.
{\small\begin{verbatim}
    $q = @B1:p - $el;      # (Electronic) momentum transfer
    Q2 = - $q^2;
    x = Q2/(2*$q*@B2:p);   # Bjorken variables
    y = Q2/(x*S);

\end{verbatim}}

Finally the same naive method is used to determine the single
outgoing quark momentum.
{\small\begin{verbatim}
    for (@p in QUARKS)     # Collect the outgoing quark (= jet).
      $jet = @p:p;
  }

\end{verbatim}}

\subsubsection{Event processing}
\label{sec:example:processing}

After having determined the Bjorken variables in the last section,
we can now use them for kinematical cuts.  The next action will only
be executed for those events that satisfy $x_{min.} \le x \le x_{max.}$
and $Q^2 \ge Q^2_{min.}$.  Note that (unlike other programming
languages) comparisons can be combined to intervalls in a natural
way.  See section \ref{sec:operators} for more details on comparisons
and logical operators.
{\small\begin{verbatim}
(x_min <= x <= x_max) && (Q2_min <= Q2)
  {
\end{verbatim}}

The first histogram will represent the total
cross section inside the kinematical cuts as a function of the
photon multiplicity and the infrared cut off in the photon energy.
\hepawk/ version \Version/ does not have a \.{for} loop similar
to \.C (\.{for} is at the moment only used for looping over
particles), thus we implement the loop over the cut offs with a
\.{while} loop.
{\small\begin{verbatim}
    E_cut = 0.0;
    while (E_cut <= E_cut_max)
      {
\end{verbatim}}

For each cut off energy we now loop over all photons and select those
that satisfy the energy cut and the angular cuts.
{\small\begin{verbatim}
        i = 0; # multiplicity counter
        for (@p in PHOTONS)
          if (@p:p:E >= E_cut)
            {
\end{verbatim}}

A photon is counted if it falls either in the forward or backward
luminosity
monitor or in the main detector.  Furthermore it has to be sufficiently
separated from the outgoing electron and quark jet.
{\small\begin{verbatim}
              theta = angle (@p:p, @B1:p);
              if ((th_min_lumi <= theta <= th_max_lumi         # bwd
                   || th_min_lumi <= PI - theta <= th_max_lumi # fwd
                   || th_min_main <= theta <= th_max_main)     # main
                  && angle (@p:p, $quark) >= th_sep_jet
                  && angle (@p:p, $el) >= th_sep_em)
                i++;
            }
\end{verbatim}}

Before reiterating the \.{while} loop,
we fill the bin in the histogram corresponding to the measured
multiplicity and the applied energy cut.
{\small\begin{verbatim}
        fill (h_m, i, E_cut + 0.01);
        E_cut = E_cut + E_cut_step;
      }

\end{verbatim}}

The other histograms record the energy spectrum of the hardest and
second hardest photon in a multiphoton event.  This is implemented
by looping over all existing photons satisfying a certain energy
cut and satisfying the same angular and separation cuts as above.
{\small\begin{verbatim}
    n = 0; $p_hard = $p_soft = $NULL;
    for (@p in PHOTONS)
      {
        if (@p:p:E >= E_min)
          {
            theta = angle (@p:p, @B1:p);
            if ((th_min_lumi <= theta <= th_max_lumi         # bw lumi
                 || th_min_lumi <= PI - theta <= th_max_lumi # fw lumi
                 || th_min_main <= theta <= th_max_main)     # main
                && (angle (@p:p, $quark) >= th_sep_jet)
                && (angle (@p:p, $el) >= th_sep_em))
              {
\end{verbatim}}

If the current photon is harder than the second
hardest photon seen in the
event so far, update the variables holding
the momenta of these two photons.
{\small\begin{verbatim}
                if (@p:p:E > $p_hard:E)       # the hardest photon yet
                  {
                    $p_soft = $p_hard; $p_hard = @p:p;
                  }
                else if (@p:p:E > $p_soft:E)  # the second hardest
                  $p_soft = @p:p;
                n++;
              }
          }
      }

\end{verbatim}}

Once we have decided which are the hardest photons in this event,
we can fill the corresponding histograms.  Furthermore a counter
is incremented which will later be used to determine the accumulated
cross section.
{\small\begin{verbatim}
    if (n == 0)
      incut_0++;     # count this event
    else if (n == 1)
      {
        incut_1++;   # count this event
        fill (h_E, $p_hard:E);
      }
    else
      {
        incut_2++;
        fill (h_E1, $p_hard:E); fill (h_E2, $p_soft:E);
        fill (h_E1_E2, $p_hard:E, $p_soft:E);
      }
  }

\end{verbatim}}

\subsubsection{The {\tt END} action}
\label{sec:example:END}

The \.{END} action will be executed after the last event has
been processed.  Typically it is used to print results and to
write histograms to a permanent storage medium.
Multiple instances of the \.{END} action
will be concatenated.

This \.{END} action starts by printing the integrated cross sections
for zero, one, and multi photon events.  These are constructed from the
counters incremented earlier, the total number of scanned events,
and the total cross section.  The latter two variables are provided
by the Monte Carlo in \hepevt/.
{\small\begin{verbatim}
END
  {
    printf ("null photon events: %d, cross section: %gmb\n",
            incut_0, (incut_0/NEVENT) * XSECT);
    printf ("one photon events: %d, cross section: %gmb\n",
            incut_1, (incut_1/NEVENT) * XSECT);
    printf ("two photon events: %d, cross section: %gmb\n",
            incut_2, (incut_2/NEVENT) * XSECT);
\end{verbatim}}
Now a line printer style plot of the histograms is printed out.
{\small\begin{verbatim}
    printf ("\nHISTOGRAMS:\n");
    printf ("***********\n\n");
    plot ();
\end{verbatim}}
And finally the contents of the histograms is normalized to
picobarns per bin and written to the file \.{"PAW"}.
{\small\begin{verbatim}
    scale (1.0e9 * XSECT/NEVENT);  # normalize to picobarns
    save ("PAW");
    printf ("\ndone.\n");
  }
\end{verbatim}}


\section{\f77/ Interface}
\label{sec:f77}

\hepawk/ reads {\em all\/} information from the standard \hepevt/ common
block \cite{hepevt}.
If the Monte Carlo writes its output into \hepevt/,
the \f77/ interface is as simple as the example in figure
\ref{ex:f77}.  After the event generator (\.{gener}) has filled the
common block, \hepawk/ is called with the argument \.{'scan'} to
analyze it.  An explicit initialization call is {\em not\/} necessary,
but useful to catch syntax errors in the script before costly
computations by the event generator have been performed.

In addition to the standard described in \cite{hepevt}, \hepawk/
requires the incoming particles to be stored in \hepevt/ too.  Following
the \.{HERWIG} convention \cite{HERWIG}, they must have the status
codes 101 and 102 respectively.

\begin{example}{\f77/ interface}{ex:f77}
\.{* mctest.f}                                                      \\
\>\>\> \.{...}                                                      \\
\>\>\> \.{call hepawk ('init')}         \C compile the script       \\
\>\>\> \.{...}                                                      \\
\>\>\> \.{do 10 n = 1, nevent}                                      \\
\>\>\> \>\> \.{call gener ()}           \C generate an event        \\
\>\>\> \>\> \.{call hepawk ('scan')}    \C analyze the event        \\
\.{10} \>\>\> \.{continue}                                          \\
\>\>\> \.{...}
\end{example}

In order to avoid possible clashes in the \f77/ name space,
all names of external symbols (functions, subroutines, and
common blocks)
are of the form \.{`hk....'} or \.{`hep...'}.  An exception
of this rule are
the external symbols of the parser, with have been given the
traditional \.{`yy...'} names.

\hepawk/ version \Version/ relies on the standard \hepevt/
common block and the HBOOK\cite{BL87} histogramming package.
It should be stressed, however, that
both are accessed through a set of interface subroutines
(\.{`hep...'} for \hepevt/ and \.{`hkh...'} for HBOOK), such that
using a different common block for passing the event information
and using a different histogramming package can be implemented
straightforwardly.


\section{Implementation}
\label{sec:implementation}

In this section we discuss the implementation of \hepawk/.
This discussion is rather detailed in order to show how similar
interpreters could be implemented along the same lines.

A straightforward interpreter which would translate the script
over and over again
for each event scanned is of no use for applications in high energy
physics, because of performance considerations.  Therefore \hepawk/
first compiles the script into code for a stack machine, which is in
turn executed for each event.  Parsing is done by a LALR(1)
parser (see e.g.~\cite{AU77}), generated
by a customized version of Bison\cite{DS88} which has been modified to
produce \f77/ code.
Bison has been chosen, because it is available as freely
redistributable
source code and we had to make some changes in order to support \f77/
parsers.  Bison is compatible to the parser generator
YACC \cite{AJU75}, which is standard in UNIX environments.

The following subsections describe in detail
the implementation of the three
main components of \hepawk/: lexical analysis, parser, and stack
machine.


\subsection{Lexical Analysis}
\label{sec:lex}

Lexical analysis in \hepawk/ is very simple.  The lexical analyzer
splits
the input into tokens and passes them on to the parser for syntactical
analysis.  Comments are ignored and multiple whitespace (blanks,
tabs and
newlines) is converted to single blanks (but the location of the tokens
is remembered for reporting errors).

Lexical analysis is further simplified by using lexical conventions
for the different types of data.  All variables denoting a four vector
have to start with a \.{\$} and all variables denoting a particle
have to start with a \.{@}.
Another approach would have been to require
the declaration of data types (e.g. \.{vector q; particle p;}, etc.),
but it is not desirable to require this overhead in \hepawk/ scripts.


\subsection{Syntactical Analysis}
\label{sec:parse}

As already mentioned, the syntactical analysis is performed by an
automatically generated, table driven LALR(1) parser.  This approach
has
several important advantages over constructing a parser from scratch:

\begin{itemize}
  \item{} An automatically generated parser recognizes all valid
      sentences in the specified grammar and only those.  It is far from
      trivial to achieve this by hand coding the parser.
  \item{} Extending the grammar or changing parts of it can be done very
      easily.  With a hand coded parser, a simple change might force the
      programmer to redesign the parser completely.
  \item{} Table driven LALR(1) parsers do {\em not\/} require recursive
      subroutine calls to parse recursive expression grammars (this
      point is relevant to standard conforming \f77/ implementations).
  \item{} Error handling is comparatively simple for LALR(1) parsers.
  \item{} A LALR(1) parser is reasonably fast.
\end{itemize}

We will now illustrate the use of a parser generator by describing the
implementation of a particular language element in detail.  In computer
science our compilation method is known as ``syntax directed
translation'' (see e.g.~\cite{AU77} for a general account).

Figure \ref{fig:while} shows a slightly simplified excerpt from the
Bison grammar for \hepawk/ in which the \.{while} control statement is
implemented (see appendix \ref{sec:grammar} for the full grammar).

\begin{figure}[htb]
\begin{tabbing}
statement \= \qquad \=: $\ldots$ \\
  \>\> $\vert$ \= \.{"while"} \{ \C recognize reserved word \\
  \> \.{call hkcpos (\$\$)}      \C remember current position \\
\} \\
  \>\> \> \.{"("} logical\_expr \.{")"}  \{ \C parse condition \\
  \> \.{call hkcasm ('JMPF')}               \C jump if false \\
  \> \.{call hkcdum (\$\$)}      \C address will be filled in below \\
\} \\
  \>\> \> statement \{           \C parse the loop body \\
  \> \.{call hkcasm ('JMP ')}    \C unconditionally jump back \\
  \> \.{call hkcins (\$2)}       \C to the condition \\
  \> \.{call hkcbck (\$6)}       \C fill in the exit address \\
\} \\
  \>\> $\vert$ $\ldots$
\end{tabbing}
\caption{Compiling \.{while} loops}
\label{fig:while}
\end{figure}

If we disregard the actions delimited by braces, it tells the parser
generator that a statement can be formed by the reserved word
\.{"while"}, a logical expression in parenthesis and a statement.

After recognizing the reserved word \.{"while"}, the first action
instructs the parser to execute the \f77/ statement
\.{call hkcpos (\$\$)}, in which the parser generator will replace the
symbol \.{\$\$} by a reference to the top of the parser's stack.  The
subroutine \.{hkcpos} will store there
the address of the next instruction in
the compiled program for later reference by the jump instruction at
the end of the \.{while} loop.  Now the parser will continue to parse the
logical expression forming the condition of the \.{while} loop.  During
this parse it generates code to evaluate this condition, leaving the
result on the top of the run-time stack.

The second action instructs the parser to generate a conditional jump
instruction (\.{call hkcasm ('JMPF')}) which will jump out of the loop if
the condition was false (i.e.~if the value on top of the run-time stack is
false).  A place holder is generated for the address to jump to and its
address is remembered on the parser's stack (\.{call hkcdum (\$\$)}), so
we can fill it in later.  This is necessary, because we do not yet how
long the following statement (which is usually a sequence of statements)
will be.

After parsing the statement and generating the corresponding code, the
third action terminates the loop.  The parser replaces the symbol
\.{\$2} by a reference to the stack location where the first action
(the second component of the production under consideration) has
stored the entry address of the \.{while} loop.  The \f77/ statement
\.{call hkcbck (\$6)} finally ``backpatches'' the unresolved address
in the second action (the sixth component).


\subsection{Stack Machine}
\label{sec:stack}

The stack machine executes the ``machine code'' stored by the parser
in an array of integers.  Evaluation of expressions is performed
by straightforward
manipulations of operands in a sufficiently large run time stack.

\begin{figure}[htb]
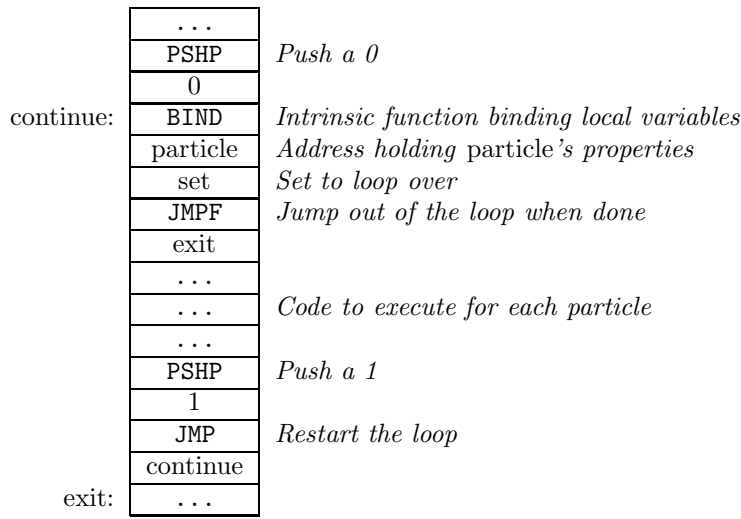

\begin{tabular}{r|c|l}
\cline{2-2}
          & \.{...}     &                               \\ \cline{2-2}
          & \.{PSHP}    & \,{Push a 0}                  \\ \cline{2-2}
          & 0           &                               \\ \cline{2-2}
continue: & \.{BIND}    & \,{Intrinsic function binding
                           local variables}             \\ \cline{2-2}
          & particle    & \,{Address holding} particle\,{'s
                           properties}          \\ \cline{2-2}
          & set         & \,{Set to loop over}          \\ \cline{2-2}
          & \.{JMPF}    & \,{Jump out of the loop
                           when done}                   \\ \cline{2-2}
          & exit        &                               \\ \cline{2-2}
          & \.{...}     &                               \\ \cline{2-2}
          & \.{...}     & \,{Code to execute for each
                           particle}                    \\ \cline{2-2}
          & \.{...}     &                               \\ \cline{2-2}
          & \.{PSHP}    & \,{Push a 1}                  \\ \cline{2-2}
          & 1           &                               \\ \cline{2-2}
          & \.{JMP}     & \,{Restart the loop}          \\ \cline{2-2}
          & continue    &                               \\ \cline{2-2}
   exit:  & \.{...}     &                               \\ \cline{2-2}
\end{tabular}
\caption{Implementation of \.{for} loops}
\label{fig:for-loop}
\end{figure}

As an example of the code for the stack machine we will discuss the
\.{for} loop which is used to loop over all particles in a given
set.  Figure \ref{fig:for-loop} shows the actual code for a \.{for}
loop.  Before entering the \.{for} loop, a 0 is pushed on the stack.
If the \.{BIND} instruction finds a 0 on the stack, it fetches the
{\em first\/} particle belonging to the desired set from \.{/hepevt/}
and stores it at an address where the body of the loop can access it.
If there are no more particles from the set, \.{BIND} pushes a
\.{FALSE} and the next instruction jumps out of the loop.  After the
body has been executed, a 1 is pushed on the stack causing the
next \.{BIND} to fetch the {\em next\/} particle.

\subsection{Timings}
\label{sec:timings}

The time spent in the compilation step is certainly negligible and
does not call for
further optimizations.  Most of this time is consumed
by the lexical analyzer and can not be reduced portably anyway.  A
typical 180 lines \hepawk/ script takes about 140ms to compile on a
IBM RS/6000 (Model 520).

On the other hand, the time spent executing the code is not
negligible.  Measurements on a IBM RS/6000 have shown that the
execution time of the equivalent straight \f77/ program will roughly be
multiplied by a factor of three.  The aforementioned 180 lines
\hepawk/ script takes about 0.6 msec to execute, where 0.2 msec are
due to scanning of the \hepevt/ common blocks and to library calls
(HBOOK).  Since the equivalent \f77/ program will have to perform the
same tasks, the remaining 0.4 msec can serve as a conservative
estimate on \hepawk/'s execution time.

Future versions will take advantage of some optimizations and should
be able to reduce the computational overhead to a factor of two.


\section{Future Developments}
\label{sec:future}

\hepawk/ has been developed in the context of a specific
project: the construction of a parton level
Monte Carlo event generator
for HERA physics.  Nevertheless it has been designed to meet more
general requirements and should be easily adapted to different
applications.  Indeed, since the release of version 1.0 \hepawk/ has
been be a useful tool in the development of the \KROWIG/ Monte Carlo
\cite{Ohl92b}, proving that it can deal conveniently and efficiently
with hadronic final states.

Amongst possible future developments for \hepawk/ are:
\begin{itemize}
\item{} User definable functions (subroutines):  these would allow to
  group commonly used kinematical calculations into a subroutine
  library.
\item{} Arrays:  most likely these will be implemented as
  associative arrays, and the \.{for} statement will allow to loop
  over all elements.  In connection with user definable functions it
  will for example be possible to sort these arrays according to some
  prescription.
\item{} More physics:  \hepawk/'s syntax can be extended to admit jet
  finding algorithms, etc.
\item{} Optimization:  for the sake of simplicity, some language
  elements are compiled sub-optimally in \hepawk/ version \Version/.
  Although there shall be no dramatic increase in running time, some
  improvements are possible,  especially for complicated logical
  expressions.
\end{itemize}


\section{Conclusions}
\label{sec:concl}

We have described the implementation of \hepawk/ version \Version/,
a programming language for scanning data structures in high energy
physics.  Implemented in \f77/, this language is completely portable
and can be used with any Monte Carlo event generator which writes
standard event records.


\section*{Acknowledgements}

I want to thank my collaborators in the \KRONOS/ project, which
triggered the development of \hepawk/.  Special thanks to Harald
Anlauf for valuable suggestions and patience in using unfinished
versions of \hepawk/.



\appendix
\section{Reference Manual}
\label{sec:ref-man}

This appendix serves as a reference manual for \hepawk/ version
\Version/.


\subsection{Input files}
\label{sec:files}

The input files are in free format (record separators are ignored apart
from terminating comments).  But to allow identical operation on all
computers used in high energy physics, the maximum length of input
lines has been restricted to 72.


\subsection{Data types}
\label{sec:data-types}

\begin{description}
    \Item{x\_elec\_1}{}
      Scalar variable.
    \Item{\$\,{p}}{}
      \,{p} is a four-vector.
    \Item{:1, :2, :3, :4, :0, :x, :y, :z, :E, :t}{}
      Extract a component from the preceding four-vector (\.{:0, :4,
      :E, :t} are different names for the time like component,
      \.{:3, :z} are different names for one of the beam directions.
      For example \.{\$p:E} and \.{\$p:0}
      both refer to the energy component of the
      four vector \.{\$p}.
    \Item{@\,{p}}{}
      \,{p} is a particle (with four-momentum, production vertex,
      and spin in the fashion of \.{/hepevt/}\cite{hepevt}).
    \Item{:id, :p, :v, :s}{}
      Extract the PDG code\cite{PDG90}, the four-momentum, vertex, and
      spin vector from the preceding particle.
      For example \.{@p:p} and \.{@p:p:E}
      refer to the four momentum and the energy of the particle \.{@p}.
\end{description}


\subsection{Statements}
\label{sec:statements}

\begin{description}
    \Item{BEGIN, END}{}
      Special tokens, indicating that the following statement is
      executed only during initialization and clean up.
    \Item{\{, \}}{}
      Start and terminate a group of statements.
    \Item{;}{}
      Terminates a statement.  (\hepawk/ does \,{not} recognize
      newline as a statement terminator!)
\end{description}


\subsection{Control Structures}
\label{sec:flow-control}

\begin{description}
    \Item{if (\,{condition}) \,{statement}}{}
      Execute \,{statement} if \,{condition} is true.
    \Item{if (\,{condition}) \,{statement\_1} else \,{statement\_2}}{}
      Execute \,{statement\_1} if \,{condition} is true, if not
      execute \,{statement\_2}.
    \Item{while (\,{condition}) \,{statement}}{}
      Repeatedly execute \,{statement} while \,{condition} is true.
    \Item{for (@\,{particle} in \,{set}) \,{statement}}{}
      Execute \,{statement} for each \,{particle} in \,{set}.
      Available \,{set}s are: \.{QUARKS}, \.{LEPTONS}, \.{PHOTONS},
      \.{HADRONS}, \.{ELECTRONS}.
    \Item{next}{}
      Terminate execution of the script here and wait for the next
      record.
    \Item{break}{}
      Jump out of the innermost enclosing \.{for} or \.{while} loop.
    \Item{continue}{}
      Restart the innermost enclosing \.{for} or \.{while} loop.
\end{description}


\subsection{Operators}
\label{sec:operators}

\begin{description}
    \Item{=}{}
      Binary arithmetical assignment operator.
    \Item{-}{}
      Unary arithmetical operator: negative ($-$).
    \Item{+, -, *, /, \^{}}{}
      Binary arithmetical operators: sum ($+$), difference ($-$),
      product ($*$), quotient ($/$), and power ($x^y$), resp.~cross
      product.  Products of
      four-vectors are either (lorentzian) inner, (three dimensional)
      cross or scalar product.
      Divison of four-vectors is availabe as right scalar division.
      division is \,{not} available for four-vectors.  The power operator
      for four-vectors is defined as $(v)^a \equiv \mbox{sign} (v\cdot v)
      \vert v\cdot v \vert^{a/2}$.  Note that $v^4 \not= (v^2)^2$ for
      spacelike $v$.
    \Item{+=, -=, *=, /=}{}
      Binary arithmetical operators with assignment (i.e. \.{a += b}
      $\equiv$ \.{a = a + b}).  These are at the moment not more
      efficient than the long form, just more concise (this may change
      in future releases).
    \Item{++, --}{}
      Unary increment and decrement operator.  Like C we distinguish
      post increment from pre increment (resp. decrement).  \.{x++} has
      the same effect as \.{x = x + 1}, but is more concise and
      slightly more efficient.
    \Item{!}{}
      Unary logical operator: not ($\neq$).
    \Item{\&\&, ||}{}
      Binary logical operators: and ($\wedge$), or ($\vee$).
    \Item{==, !=, <, <=, >, >=}{}
      Binary arithmetical comparison operators: equal ($=$),
      not equal ($\neq$), less than ($<$), less than or equal to ($\leq$),
      greater than ($>$), greater than or equal to ($\geq$).
      Unlike most other programming languages these operators can be
      combined to form intervals, thus \.{0.01 < x < 0.5} is a well
      defined condition in \hepawk/.  This expression is identical to
      \.{0.01 < x \&\& x < 0.5}, but \.{x} is evaluated only once.
      Furthermore even \.{0.01 < x\_1 <= x\_2 < 0.5} is legal and has
      the obvious semantics.  Please note that version \Version/ of
      \hepawk/ does {\em not\/} optimize the evaluation of logical
      expressions in the way C does it.  This may change in a future
      version.
    \Item{/|}{}
      $\Lambda$ operator, implementing a
      Lorentz boost: \.{\$q = \$p /| \$beta} corresponds to
      \begin{eqnarray*}
        \gamma & = & \frac{1}{\sqrt{1-\vec\beta^2}} \\
        q_0 & = & \gamma \left(p_0 - \vec\beta\vec p\right) \\
        \vec q & = & \vec p + \vec\beta\left( (\gamma - 1)
                       \frac{\vec\beta\vec p}{\vec\beta^2}
                          - \gamma p_0 \right)
      \end{eqnarray*}
\end{description}


\subsection{Builtin Constants}
\label{sec:constants}

\begin{description}
    \Item{ALPHA}{}
      Fine structure constant $\alpha_{{\rm QED}} = 1/137.03599$.
    \Item{DEG}{}
      Degrees per radians: $180/\pi$, this is useful if you prefer
      angular cuts in degrees: \.{10/DEG < angle (\$p, @B1:p)}.
    \Item{GAMMA}{}
      Euler's $\gamma = 0.577215664901532861$.
    \Item{HBARC}{}
      Conversion factor $\hbar c = 197.3271 \hbox{ MeV fm}$.
    \Item{HBC2}{}
      Conversion factor $(\hbar c)^2 = 0.3893797 \mbox{GeV}^2 \mbox{mbarn}$.
    \Item{PI}{}
      The number $\pi = 3.141592653589793238$.
    \Item{NULL, \$NULL, @NULL}{}
      The number $0$, the null four-vector $(0; 0, 0, 0)$, and a
      particle with all components $0$.
    \Item{\$E1, \$E2, \$E3, \$E0}{}
      Unit four-vectors $(0; 1, 0, 0)$, $(0; 0, 1, 0)$, $(0; 0, 0,
      1)$, $(1; 0, 0, 0)$.  Available aliases are \.{\$EX, \$EY,
      \$EZ, \$ET, \$EE, \$E4}.
    \Item{\_pdg\_electron, \_pdg\_proton, ...}{}
      Symbolic names for some PDG \cite{PDG90} particle codes.  Also
      available are (to be prefixed by \.{\_pdg\_}):
      \.{down}, \.{up}, \.{strange}, \.{charm},
      \.{bottom}, \.{top}, \.{electron}, \.{nu\_e},
      \.{muon}, \.{nu\_m}, \.{tau}, \.{nu\_t},
      \.{gluon}, \.{gluon\_}, \.{photon}, \.{Z0},
      \.{W\_plus}, \.{higgs0}, \.{higgsp}, \.{pi0},
      \.{eta}, \.{K\_plus}, \.{K0}, \.{K0\_short},
      \.{K0\_long}, \.{proton}, \.{neutron}.
\end{description}


\subsection{Automatic variables}
\label{sec:autovar}

These variables are initialized automatically for each record.
\hepawk/ follows the \.{HERWIG} convention \cite{HERWIG}    and recognizes
the first two entries with status code 101 and 102 as beam particles.
The user is responsible for providing these.

\begin{description}
    \Item{@B1, @B2}{}
      The two incoming (beam) particles.
    \Item{NEVENT}{}
      The current event number.  In \.{END} actions this is the total
      number of events.
    \Item{XSECT}{}
      The total generated cross section in mb.  Non-zero in \.{END}
      actions only.
    \Item{ERROR}{}
      The Monte Carlo error (in mb), as estimated by the calling Monte
      Carlo event generator.  Non-zero in \.{END} actions only.
      {\it (NB: this variable is only defined, iff the calling Monte
      Carlo stores the error in \.{phep(1,2)}, which is not required
      by the standard).}
    \Item{MC}{}
      Monte Carlo identification number (as in \.{/hepevt/}).
    \Item{REV}{}
      Monte Carlo Version formatted as string.
    \Item{RUN}{}
      Run identification number.
    \Item{DATE}{}
      Date of run formatted as string.
\end{description}


\subsection{Builtin Functions}
\label{sec:functions}


\subsubsection{Arithmetic}
\label{sec:arithmetic}

\begin{description}
    \Item{sin ($x$), cos ($x$), tan ($x$)}{}
      Returns the sine, cosine, tangent of $x$.
    \Item{asin ($x$), acos ($x$), atan ($x$)}{}
      Returns the inverse sine, cosine, tangent of $x$.
    \Item{atan2 ($x$, $y$)}{}
      Returns the inverse tangent of $y/x$.
    \Item{abs ($x$)}{}
      Returns the absolute value $|x|$ of $x$.
    \Item{sqrt ($x$)}{}
      Returns the square root of $x$.
    \Item{ssqrt ($x$)}{}
      Returns $\mbox{sign}(x)\cdot\sqrt x$.  This is useful in
      \.{ssqrt(\$q*\$q)}, if $q$ might be spacelike.
    \Item{log ($x$), log10 ($x$), exp ($x$)}{}
      Returns the logarithms to base $e$ and $10$ and the exponential of $x$.
    \Item{angle (\$$p$, \$$q$)}{}
      Returns the angle between the spatial parts of the four-vectors
      $p$ and $q$.
      \begin{equation}
        \mbox{\.{angle}} (p, q)
           = \mbox{acos} \left( \frac{\vec p \cdot \vec q}
                                     {|\vec p| \cdot |\vec q|} \right)
        \label{op:angle}
      \end{equation}
    \Item{rap (\$$p$), prap (\$$p$), pt (\$$p$)}{}
      Rapidity, pseudo rapidity, and transverse momentum of the
      four-vector $p$.
      \begin{eqnarray}
        \mbox{\.{rap}} (p) & = & y(p) = \frac{1}{2}
           \ln \left( \frac{p_0 + p_3}{p_0 - p_3} \right) \\
        \mbox{\.{prap}} (p) & = & \eta(p) = - \ln \left( \tan
           \left( \frac{1}{2} \angle(\vec p, \vec e_z) \right) \right) \\
        \mbox{\.{pt}} (p) & = & \sqrt{p_1^2 + p_2^2}
        \label{op:pt}
      \end{eqnarray}
    \Item{dist (\$$p$, \$$q$), pdist (\$$p$, \$$q$)}{}
      Distance of the four-vectors $p$ and $q$
      in rapidity space (resp. pseudo rapidity space).
      \begin{eqnarray}
        \mbox{\.{dist}} (p, q)
            & = & \sqrt{ (\Delta y)^2 + (\Delta\phi)^2 } \\
        \mbox{\.{pdist}} (p, q)
            & = & \sqrt{ (\Delta\eta)^2 + (\Delta\phi)^2 }
        \label{op:dist}
      \end{eqnarray}
\end{description}


\subsubsection{Output}
\label{sec:output}

\begin{description}
\Item{printf (\,{format}, \,{arg1}, \,{arg2}, \,{arg3}, $\ldots$)}{}
  Print the arguments according to the format specified in \,{format}.
  The following formats and escapes are implemented in \hepawk/ version
  \Version/:
  \begin{itemize}
  \item{} \.{\%d}: integer.
  \item{} \.{\%e, \%f, \%g}: floating point number in exponential,
    fixed, or mixed representation.
  \item{} \.{\%s}: string.
  \item{} \.{\%v, \%u, \%w}: floating point four-vector in the format
    $(v_0; v_1, v_2, v_3)$ in exponential,
    fixed, or mixed representation.
  \item{} \.{$\backslash$n}: insert newline.
  \end{itemize}
  It is likely that future versions of \hepawk/ will allow to specify
  field widths in the format string.
\end{description}


\subsubsection{Miscellaneous}
\label{sec:misc}

\begin{description}
    \Item{rand ()}{}
      Return a on the intervall $[ 0, 1]$
      uniformly distributed random number, generated by the
      algorithm of \cite{MZT90}.
    \Item{rseed (\,{ijseed}, \,{klseed})}{}
      Reinitialize the random number generator with the seeds
      \,{ijseed} (mod 31329) and \,{klseed} (mod 30082).
    \Item{unidev ($\mu$, $\sigma$)}{}
      Return a on the intervall $[ \mu-\sigma, \mu+\sigma]$
      uniformly distributed random number.
      Note that $\sigma^2$ is {\em not\/} the second moment of
      the generated distribution.
    \Item{gauss ($\mu$, $\sigma$)}{}
      Return a gaussian distributed random number with mean $\mu$
      and standard deviation $\sigma$.  See figure \ref{ex:charge}
      for a sample application.
    \Item{charge ($i$)}{}
      Calculate the charge of a particle from its PDG\cite{PDG90} code
      $i$, see figure \ref{ex:charge} for a sample application.
    \Item{dump (\,{flags})}{}
      Dump the current event completely.  This can be useful for
      debugging.  The optional string \,{flags} can be used to
      suppress some output: \.{p}: momenta, \.{v}: production vertex,
      \.{s}: spin, \.{m}: mothers, \.{d}: daughters, and \.{r}:
      reserved status codes.
    \Item{exit ()}{}
      Terminate the program bye executing a \f77/ {\tt STOP} instruction.
\end{description}

\begin{example}{Charged particles}{ex:charge}
\.{\# charge.hepawk}                             \\
\.{...}                                         \\
\LB                                             \\
\.{for (@p in HADRONS)}\+    \C select hadrons  \\
\.{if (rand () < 0.9 \&\& charge (@p:id))}
                             \C select charge   \\
\LB                   \C  with 90\% efficiency  \\
\.{energy = @p:p:E;}         \C naive energy    \\
\.{energy = gauss (energy, 0.16 * sqrt (energy));}
                                \C calorimeter  \\
\.{fill (charged\_hadrons, energy);} \C detected\\
\RB\-                                           \\
\RB                                             \\
\.{...}
\end{example}


\subsubsection{Histogramming}
\label{sec:histogramming}

\begin{description}
    \Item{book1 ($n$, \,{title}, $i$, $x_{\min}$, $x_{\max}$)}{}
      Book a one-dimensional histogram with identifier $n$, title
      \,{title}, and  $i$ channels in the interval $[x_{\min}, x_{\max}]$.
      Only the identifier is required, for the other parameters the
      following defaults are provided: \,{title} $=$ \.{" "}, $i =
      100$, $[x_{\min}, x_{\max}] =$ autoscaling.
      This and the following booking functions return the identifier
      of the booked histogram.  This is useful for the special
      argument $n\le0$, in which case a unique identifier will be
      created by \hepawk/.
    \Item{book2 ($n$, \,{title}, $i$, $x_{\min}$, $x_{\max}$, $j$,
      $y_{\min}$, $y_{\max}$)}{}
      Book a two-dimensional histogram with identifier $n$, title
      \,{title}, $i$ channels in the interval $[x_{\min}, x_{\max}]$,
      and  $j$ channels in the interval $[y_{\min}, y_{\max}]$.
      All parameters are required.
    \Item{bookl ($n$, \,{title}, $i$, $x_{\min}$, $x_{\max}$)}{}
      Book a one-dimensional histogram with identifier $n$, title
      \,{title}, and  $i$ channels in the interval $[x_{\min},x_{\max}]$.
      These channels will be {\em logarithmically\/} equidistant.
      All parameters are required.
    \Item{bookp ($n$)}{}
      Book the projections for the two-dimensional histogram $n$.
      ($n = 0$ applies to all two-dimensional histograms booked so
      far.)
    \Item{fill ($n$, $x$, $y$)}{}
      Fill the point $(x,y)$ into the histogram $n$.  If $y$ is not
      supplied, it defaults to $0$.
    \Item{plot ($n$)}{}
      Plot the histogram $n$ in lineprinter format.  If $n$ is not
      supplied, it defaults to $0$, which means plot all booked
      histograms.
    \Item{scale ($n$, $\mu$)}{}
      Rescale the contents of histogram number $n$ by the factor
      $\mu$.  This function is convenient for normalizing histograms
      to total cross sections,
      since the latter usually are available only after
      the run.  The abbreviation \.{scale ($\mu$)} is equivalent to
      \.{scale ($0$, $\mu$)}, which means rescale all histograms.
    \Item{arith ($n$, $i_1$, \,{op}, $i_2$, $x_1$, $x_2$)}{}
      Perform arithmetic on the histograms $i_1$ and $i_2$.
      If $\chi(k;l)$ denotes the contents of channel $l$ of histogram
      $k$, then we have
      \begin{equation}
        \chi(n;l) = x_1 \chi(i_1;l) \mbox{\,{op}} x_2 \chi(i_2;l)
      \end{equation}
      and \,{op} is one of \.{+}, \.{-}, \.{*}, or \.{/}.
      Iff $n=0$ a new histogram is allocated and its number is returned.
    \Item{copy ($n$, $i$, \,{title})}{}
      Copy histogram $i$ to histogram $n$, optionally changing the title
      to \,{title}.  Iff $n=0$ a new histogram is allocated and its
      number is returned.
    \Item{delete ($n$)}{}
      Delete histogram $n$, with $n=0$ meaning all histograms.
    \Item{save (\,{filename})}{}
      Save the current directory in the permanent file \,{filename}.
    \Item{open (\,{filename}, \,{directory}, \,{mode})}{}
      Open the \.{RZ} file \,{filename} and connect it to the \.{RZ}
      directory \,{directory}.
    \Item{close (\,{directory})}{}
      Close the \.{RZ} file connected to the \.{RZ} directory
      \,{directory}.
    \Item{read ($n$, \,{cycle}, \,{offset})}{}
      Read the cycle \,{cycle} of histogram $n$ from disk into
      histogram $n + \mbox{\,{offset}}$.  Default value for \,{offset}
      is $0$ and $9999$ for \,{cycle}.  (\,{cycle} and \,{offset}
      might not be supported with other histogramming packages than
      \.{HBOOK}.)
    \Item{write ($n$)}{}
      Write the histogram $n$ to disk.
    \Item{mkdir (\,{dir})}{}
      Create the subdirectory \,{dir}.
    \Item{cd (\,{dir})}{}
      Change the working directory to \,{dir}.
    \Item{ls (\,{dir})}{}
      List the contents of the directory \,{dir}, this defaults to the
      current working directory.
\end{description}


\subsubsection{Graphics}
\label{sec:graphics}

These experimental features were new with version 1.4 and have
disappeared with version 1.6.
\begin{description}
    \Item{display ($\phi$, $\theta$, $r$, \,{device}, \,{filename})}{}
      Do nothing.
\end{description}


\subsection{Grammar}
\label{sec:grammar}

The following is a complete description of the grammar understood by
\hepawk/ version \Version/.  This grammar has two harmless
ambiguities, which result in so called ``shift/reduce'' conflicts for
the parser.  The first ambiguity is caused by the standard ``dangling
else'' \cite{AU77}
problem.  The second ambiguity is caused by accepting real
expressions as logical expressions.  The parser cannot decide whether
it should convert the real expression in \.{if((\,{real}))} to logical
before or after parsing the parenthesis.  But since parsing the
parenthesis does not cause a semantic action, this ambiguity can be
ignored.

A complete \hepawk/ script is a -- possibly empty -- sequence of
actions, where
an action is essentially a brace delimited sequence of statements.  If
it is preceded by a logical expression, the action is executed only if
this logical expression evaluates to true.
The simplest statements are scalar or vector expressions, terminated
by a semicolon.  Furthermore statements can be combined with
conditionals and be grouped by braces.
Logical expressions can be combined by the usual operators.  An unsual
feature is the possibility to combine numerical comparisons to ranges
(e.g.~ $10^{-3} < x < 0.1$).  A single real is regarded as true if and
only if it is not zero.
The usual rules for real arithmetic apply, enriched  by pre- and
postincrement, function calls and structure members, similarily for
vector arithmetic.  Finally,
arguments to functions are a -- possibly empty -- comma separated
list.


\subsubsection{Tokens and Operators}

This is a complete list of all tokens and operators on \hepawk/ version
\Version/.  Operators are sorted according to increasing priority.

\begin{itemize}
  \item{}tokens: {\tt "if"} {\tt "else"} {\tt "while"} {\tt "for"}
     {\tt "in"} SET {\tt "next"} {\tt "break"} {\tt "continue"}
     {\tt "BEGIN"} {\tt "END"} {\tt "("} {\tt ")"} {\tt "\{"}
     {\tt "\}"} {\tt ","} {\tt ";"} SCALAR VECTOR VECTC PARTCL PARTV
     PARTS FCTN STRING
  \item{}right associative:    {\tt "="} {\tt "+="} {\tt "-="}
     {\tt "*="} {\tt "/="}
  \item{}left associative:     {\tt "||"}
  \item{}left associative:     {\tt "\&\&"}
  \item{}left associative:     RELOP
  \item{}left associative:     {\tt "+"} {\tt "-"}
  \item{}left associative:     {\tt "*"} {\tt "/"}
  \item{}left associative:     NEG {\tt "!"}
  \item{}left associative:     {\tt "/|"}
  \item{}right associative:    {\tt "\^{}"}
  \item{}left associative:     {\tt "++"} {\tt "--"}
\end{itemize}

\subsubsection{Grammar}

Here is the list of all productions in the \hepawk/ grammar
(this replaces the syntax diagrams of \cite{Ohl92a}):

\begin{tabbing}
syntactical unit$\rightarrow$\=\qquad\=\kill
script$\rightarrow$\>\>  /* empty */
      \\\>$\vert$\> script action
      \\
action$\rightarrow$\>\> beg\_or\_end {\tt "\{"} stmt\_list {\tt "\}"}
      \\\>$\vert$\> {\tt "\{"} stmt\_list {\tt "\}"}
      \\\>$\vert$\> log\_expr
           {\tt "\{"} stmt\_list {\tt "\}"}
      \\
beg\_or\_end$\rightarrow$\>\> {\tt "BEGIN"}
      \\\>$\vert$\> {\tt "END"}
      \\
stmt\_list$\rightarrow$\>\> /* empty */
      \\\>$\vert$\> stmt\_list stmt
      \\
stmt$\rightarrow$\>\>   error {\tt ";"}
      \\\>$\vert$\> {\tt ";"}
      \\\>$\vert$\> {\tt "next"} {\tt ";"}
      \\\>$\vert$\> {\tt "break"} {\tt ";"}
      \\\>$\vert$\> {\tt "continue"} {\tt ";"}
      \\\>$\vert$\> real\_expr {\tt ";"}
      \\\>$\vert$\> vector\_expr {\tt ";"}
      \\\>$\vert$\> if\_clause stmt
      \\\>$\vert$\> if\_clause stmt {\tt "else"}
           stmt
      \\\>$\vert$\> {\tt "while"}
           {\tt "("} log\_expr {\tt ")"}
           stmt
      \\\>$\vert$\> {\tt "for"}
           {\tt "("} PARTCL {\tt "in"} SET {\tt ")"}
           stmt
      \\\>$\vert$\> {\tt "\{"} stmt\_list {\tt "\}"}
      \\
if\_clause$\rightarrow$\>\> {\tt "if"} {\tt "("} log\_expr {\tt ")"}
      \\
log\_expr$\rightarrow$\>\> {\tt "("} log\_expr {\tt ")"}
      \\\>$\vert$\> log\_expr {\tt "\&\&"} log\_expr
      \\\>$\vert$\> log\_expr {\tt "||"} log\_expr
      \\\>$\vert$\> {\tt "!"} log\_expr
      \\\>$\vert$\> real\_expr
      \\\>$\vert$\> range
      \\
range$\rightarrow$\>\> real\_expr RELOP real\_expr
      \\\>$\vert$\> range RELOP real\_expr
      \\
real\_expr$\rightarrow$\>\> SCALAR
      \\\>$\vert$\> VECTOR VECTC
      \\\>$\vert$\> PARTCL PARTS
      \\\>$\vert$\> PARTCL PARTV VECTC
      \\\>$\vert$\> {\tt "++"} real\_lval
      \\\>$\vert$\> {\tt "--"} real\_lval
      \\\>$\vert$\> real\_lval {\tt "++"}
      \\\>$\vert$\> real\_lval {\tt "--"}
      \\\>$\vert$\> real\_lval {\tt "="} real\_expr
      \\\>$\vert$\> real\_lval {\tt "+="}
          real\_expr
      \\\>$\vert$\> real\_lval {\tt "-="}
          real\_expr
      \\\>$\vert$\> real\_lval {\tt "*="}
          real\_expr
      \\\>$\vert$\> real\_lval {\tt "/="}
          real\_expr
      \\\>$\vert$\> vector\_expr {\tt "*"} vector\_expr
      \\\>$\vert$\> vector\_expr {\tt "\^"} real\_expr
      \\\>$\vert$\> FCTN {\tt "("} arg\_list {\tt ")"}
      \\\>$\vert$\> FCTN {\tt "("} error {\tt ")"}
      \\\>$\vert$\> {\tt "("} real\_expr {\tt ")"}
      \\\>$\vert$\> {\tt "-"} real\_expr \%prec NEG
      \\\>$\vert$\> real\_expr {\tt "+"} real\_expr
      \\\>$\vert$\> real\_expr {\tt "-"} real\_expr
      \\\>$\vert$\> real\_expr {\tt "*"} real\_expr
      \\\>$\vert$\> real\_expr {\tt "/"} real\_expr
      \\\>$\vert$\> real\_expr {\tt "\^"} real\_expr
      \\
real\_lval$\rightarrow$\>\> SCALAR
      \\\>$\vert$\> VECTOR VECTC
      \\\>$\vert$\> PARTCL PARTS
      \\\>$\vert$\> PARTCL PARTV VECTC
      \\
vector\_expr$\rightarrow$\>\> VECTOR
      \\\>$\vert$\> PARTCL PARTV
      \\\>$\vert$\> vect\_lval {\tt "="} vector\_expr
      \\\>$\vert$\> vect\_lval {\tt "+="}
          vector\_expr
      \\\>$\vert$\> vect\_lval {\tt "-="}
          vector\_expr
      \\\>$\vert$\> vect\_lval {\tt "*="}
          real\_expr
      \\\>$\vert$\> vect\_lval {\tt "/="}
          real\_expr
      \\\>$\vert$\> {\tt "("} vector\_expr {\tt ")"}
      \\\>$\vert$\> {\tt "-"} vector\_expr \%prec NEG
      \\\>$\vert$\> real\_expr {\tt "*"} vector\_expr
      \\\>$\vert$\> vector\_expr {\tt "*"} real\_expr
      \\\>$\vert$\> vector\_expr {\tt "/"} real\_expr
      \\\>$\vert$\> vector\_expr {\tt "+"} vector\_expr
      \\\>$\vert$\> vector\_expr {\tt "-"} vector\_expr
      \\\>$\vert$\> vector\_expr {\tt "/|"} vector\_expr
      \\
vect\_lval$\rightarrow$\>\> VECTOR
      \\\>$\vert$\> PARTCL PARTV
      \\
string\_expr$\rightarrow$\>\>  STRING
      \\
arg\_list$\rightarrow$\>\> /* empty */
      \\\>$\vert$\>  arg
      \\\>$\vert$\>  arg\_list {\tt ","} arg
      \\
arg$\rightarrow$\>\>    real\_expr
      \\\>$\vert$\> vector\_expr
      \\\>$\vert$\> string\_expr
      \\\end{tabbing}


\subsection{\f77/ Interface}
\label{sec:f77-ref}

\hepawk/ is defined as a \f77/ subroutine with a single argument:

\begin{quote}\begin{tabbing}
\tab\tab\tab\kill
\.{* hepawk.f}\+\+\+                                                    \\
\.{subroutine hepawk (opcode)}                                          \\
\.{character*4 opcode}                                                  \\
\.{...}                                                                 \\
\.{end}
\end{tabbing}\end{quote}

If the argument is \.{'init'}, \hepawk/ initializes itself and compiles
the script.  If the argument is \.{'scan'} normal processing is done.

It is not necessary to initialize \hepawk/ explicitly (it will
initialize itself during the first call).  Anyway it is recommended to
compile the script {\it before\/} any expensive calculations are performed
in the calling program, in order to catch possible syntax errors.

\hepawk/ expects to find the event in the standard \hepevt/ common
block\cite{hepevt}.

All symbols exported by the \hepawk/ library are named
\.{`hk....'}, \.{`yy....'}, or \.{`hep...'}.


\section{Test Run}
\label{sec:testrun}

\subsection{Test program}

This is a simple test program for testing \hepawk/ without a full
Monte Carlo Event generator.
The program reads the data for some particles from standard input
and stores it in \hepevt/.

{\small
\begin{verbatim}
c Very simple test driver, reading an event from unit 5 with an
c arbitrary number of particles in the format:
c
c       id    status        p1        p2        p3        p0         m
c       v1        v2        v3        v0        s1        s2        s3
c
c     2212       102      0.00      0.00   -820.00    820.00      0.00
c     0.00      0.00      0.00      0.00      0.00      0.00      0.00

      program test
      implicit none

      integer hepent
      integer n, type, status
      double precision p(5), v(4), s(3)

c compile the script
      call hepawk ('init')

c trigger BEGIN action:
c   MC id:      190416
c   MC rev:       1.37
c   MC date:  01/01/91
c   Run id:       2200
c   Run date: 01/02/91
c   Run time:    12:00
      call hepeni (190416, 137, 910101, 2200, 910102, 120000)
      call hepawk ('scan')

c read the event
      call hepnew (1)
 10   continue
         read (5, *, err=999, end=999) type, status, p
         read (5, *, err=999, end=999) v, s
         n = hepent (type, status, 0, p, v, s)
      goto 10
 999  continue

c scan it
      call hepawk ('scan')

c trigger END action
c  1,000,000 events
c    3.14159 mbarn
      call hepens (1000000, 3.14159d0, 2.718281d0)
      call hepawk ('scan')

      end
\end{verbatim}
} 


\subsection{Test Input}

This is input for the above test program.
{\small
\begin{verbatim}
        11       101      0.00      0.00     30.00     30.00      0.00
      0.00      0.00      0.00      0.00      0.00      0.00      0.00
      2212       102      0.00      0.00   -820.00    820.00      0.00
      0.00      0.00      0.00      0.00      0.00      0.00      0.00
        11         1     10.00     -5.00     25.00     27.39      0.00
      0.00      0.00      0.00      0.00      0.00      0.00      0.00
        22         1     40.00     30.00     10.00     50.99      0.00
      0.00      0.00      0.00      0.00      0.00      0.00      0.00
\end{verbatim}}


\subsection{Test Script}

Some errors have been deliberately built in to test the error reporting
routines.
{\small
\begin{verbatim}
BEGIN
  {
    printf ("Welcome to the HEPAWK test suite!\n\n");

    printf ("You *should* have seen a syntax error above\n");
    printf ("complaining about a misplaced continue statement!\n\n");
    continue; # HEPAWK will complain

    printf ("The following PDG codes should match:\n");
    printf ("   Particle       PDG Code      Charge\n");
    printf ("   electron %d =   11   %f = -1.0000\n",
            _pdg_electron, charge (_pdg_electron));
    printf ("   proton   %d = 2212   %f =  1.0000\n",
            _pdg_proton, charge (_pdg_proton));
    printf ("\n");

    printf ("Expect a runtime error (division by zero) here:\n");
    1/0;
    printf ("Expect another runtime error (invalid argument) here:\n");
    log10 (-1.0);
    printf ("\n");
  }

  {
    # The following checks crucially depend on the proper input data.
    printf ("Analyzing event #%d:\n", NEVENT);
    printf ("  Beam #1:\n");
    printf ("    PDG code:        %d (= 11)\n", @B1:id);
    printf ("    momentum:        %v\n", @B1:p);
    printf ("  Beam #2:\n");
    printf ("    PDG code:        %d (= 2212)\n", @B2:id);
    printf ("    momentum:        %v\n", @B2:p);

    printf ("  Outgoing electron:\n");
    for (@e in ELECTRONS)
      {
        printf ("    momentum:        %v\n", @e:p);
        printf ("    pt:              %e (= 0.1118E+02)\n", pt (@e:p));
        printf ("    rapidity:        %e (= 0.1544E+01)\n", rap (@e:p));
        printf ("    pseudo rapidity: %e (= 0.1544E+01)\n",
                prap (@e:p));
        printf ("    angle (deg):     %e (= 0.2409E+02)\n",
                DEG * angle (@e:p, @B1:p));
      }
    printf ("  Outgoing photon:\n");
    for (@p in PHOTONS)
      {
        printf ("    momentum:        %v\n", @p:p);
        printf ("    pt:              %e (= 0.5000E+02)\n", pt (@p:p));
        printf ("    rapidity:        %e (= 0.1987E+02)\n", rap (@p:p));
        printf ("    pseudo rapidity: %e (= 0.1987E+02)\n",
                prap (@p:p));
        printf ("    angle (deg):     %e (= 0.7869E+02)\n",
                DEG * angle (@p:p, @B1:p));
      }
    printf ("  Photon-Electron subsystem:\n");
    printf ("    invariant mass:  %e (= 0.1793E+04)\n",
            (@p:p + @e:p)^2);
    printf ("    rapidity dist.:  %e (= 0.1742E+01)\n",
            dist (@p:p, @e:p));
    printf ("\n");
  }

BEGIN
  {
    printf ("That's the end of the BEGIN actions!\n\n");
  }

END
  {
    printf ("Finally the END actions:\n\n");
    if (10 > PI)
      printf ("Conditions work!\n");
    else
      printf ("Conditions are buggy!\n");

    printf ("Counting to three:");
    i = 1;
    while (i <= 3)
      printf (" %d", i++);
    printf (".\n\n");

    printf ("Summary:\n");
    printf ("number of events  : %d (= 1000000)\n", NEVENT);
    printf ("cross section (mb): %e (= 0.3142E+01)\n", XSECT);

    printf ("That's it.\n");
  }
\end{verbatim}
} 


\subsection{Sample Output}

This is the output from the above test job.

{\small
\begin{verbatim}
 hepawk: message: starting HEPAWK, Version  1.00/00, (build 911013/1931)
 hkcbrk: error:   continue outside of for or while loop, ignored
 Welcome to the HEPAWK test suite!

 You *should* have seen a syntax error above
 complaining about a misplaced continue statement!

 The following PDG codes should match:
    Particle       PDG Code      Charge
    electron         11 =   11      -1.0000 = -1.0000
    proton         2212 = 2212       1.0000 =  1.0000

 Expect a runtime error (division by zero) here:
 hkeerr: error:   division by zero (result: 0.0):
    19:     1/0;
        -----<>---------------------------------------------------------
 Expect another runtime error (invalid argument) here:
 hkeerr: error:   negative argument (using absolute value).
    21:     log10 (-1.0);
        ----<====>------------------------------------------------------

 That's the end of the BEGIN actions!

 Analyzing event #         1:
   Beam #1:
     PDG code:                11 (= 11)
     momentum:        (0.30E+02; 0.00E+00, 0.00E+00, 0.30E+02)
   Beam #2:
     PDG code:              2212 (= 2212)
     momentum:        (0.82E+03; 0.00E+00, 0.00E+00, -.82E+03)
   Outgoing electron:
     momentum:        (0.27E+02; 0.10E+02, -.50E+01, 0.25E+02)
     pt:              0.1118E+02 (= 0.1118E+02)
     rapidity:        0.1544E+01 (= 0.1544E+01)
     pseudo rapidity: 0.1544E+01 (= 0.1544E+01)
     angle (deg):     0.2409E+02 (= 0.2409E+02)
   Outgoing photon:
     momentum:        (0.51E+02; 0.40E+02, 0.30E+02, 0.10E+02)
     pt:              0.5000E+02 (= 0.5000E+02)
     rapidity:        0.1987E+00 (= 0.1987E+02)
     pseudo rapidity: 0.1987E+00 (= 0.1987E+02)
     angle (deg):     0.7869E+02 (= 0.7869E+02)
   Photon-Electron subsystem:
     invariant mass:  0.1793E+04 (= 0.1793E+04)
     rapidity dist.:  0.1742E+01 (= 0.1742E+01)

 Finally the END actions:

 Conditions work!
 Counting to three:          1          2          3.

 Summary:
 number of events  :    1000000 (= 1000000)
 cross section (mb): 0.3142E+01 (= 0.3142E+01)
 That's it.
\end{verbatim}
} 


\section{Revision History}
\label{sec:history}

\subsection*{Version 1.6, March 1995}
\begin{itemize}
  \item{} Disabled feature:
    \begin{itemize}
      \item{} Abandoned simple graphics support (identified as
        ``creeping featurism'').
    \end{itemize}
  \item{} Miscellaneous:
    \begin{itemize}
      \item{} Improved configuration and portability.
      \item{} Small bug fixes.
    \end{itemize}
\end{itemize}

\subsection*{Version 1.5, June 1994}
\begin{itemize}
  \item{} Miscellaneous:
    \begin{itemize}
      \item{} Improved configuration.
    \end{itemize}
\end{itemize}

\subsection*{Version 1.4, June 1994}
\begin{itemize}
  \item{} New feature:
    \begin{itemize}
      \item{} Simple graphics support.
    \end{itemize}
\end{itemize}

\subsection*{Version 1.3, February 1994}
\begin{itemize}
  \item{} New syntactical features:
    \begin{itemize}
      \item{} cross product \.{\^{}}.
    \end{itemize}
  \item{} Internal changes:
    \begin{itemize}
      \item{} Make the \.{'BIND'} instruction reentrant (this fixes a
        long standing (but unnoticed) bug.
    \end{itemize}
  \item{} Miscellaneous:
    \begin{itemize}
      \item{} Convert from \.{PATCHY} to \.{noweb} and \.{autoconf}.
        Put all system dependencies into a single file.
    \end{itemize}
\end{itemize}

\subsection*{Version 1.2, June 1992}
\begin{itemize}
  \item{} New syntactical features:
    \begin{itemize}
      \item{} Lorentz boosts \.{/|}.
      \item{} Assignment operators \.{+=}, \.{-=}, \.{*=}, and \.{/=}.
      \item{} Structure and vector components are now lvalues.
    \end{itemize}
  \item{} New builtin functions:
    \begin{itemize}
      \item{} \.{arith ()}
      \item{} \.{copy ()}
      \item{} \.{delete ()}
    \end{itemize}
\end{itemize}

\subsection*{Version 1.1, March 1991}
Maintenance release.
\begin{itemize}
  \item{} \.{ERROR} automatic variable.
  \item{} All calculations are done in \.{double precision}.
  \item{} Fixing a serious bug in the parser skeleton for \.{yypars
    ()}:  \f77/ is allowed to reorder statements in a conditional
    expression, this could cause an access violation in one of the
    parser tables (noticed on VAX Fortran).
  \item{} This version has no more than 19 continuation lines in a
    single statement (\f77/ standard).
  \item{} More extensive runtime error checking.
  \item{} Minor bug fixes.
\end{itemize}

\subsection*{Version 1.0, October 1991}

First official release, submitted to the {\em Computer Physics
Communication Library}.

\section{Availability}

The latest release of \hepawk/ is available by anonymous ftp from
\begin{verbatim}
  crunch.ikp.physik.th-darmstadt.de
\end{verbatim}
in the directory
\begin{verbatim}
  pub/ohl/hepawk
\end{verbatim}
or on the World Wide Web at the URL
\begin{verbatim}
  http://crunch.ikp.physik.th-darmstadt.de/monte-carlos.html#hepawk
\end{verbatim}

\section{Installation}

\subsection{UNIX Systems}

On UNIX systems, the configuration, compilation and installation
can be performed automatically according to the following familiar
sequence:
\begin{verbatim}
    $ ./configure
    $ make
    $ make test
    $ make install
\end{verbatim}

Figure~\ref{fig:configure} shows the command line options of the
\texttt{configure} script for \hepawk/ on UNIX systems.  This
\texttt{configure} script has been created by the popular GNU
Autoconf\cite{autoconf} package and should work on all UNIX variants.

\begin{figure}
{\small\begin{verbatim}
Usage: configure [options] [host]
Options: [defaults in brackets after descriptions]
Configuration:
  --cache-file=FILE       cache test results in FILE
  --help                  print this message
  --no-create             do not create output files
  --quiet, --silent       do not print `checking...' messages
  --version               print the version of autoconf that created configure
Directory and file names:
  --prefix=PREFIX         install architecture-independent files in PREFIX
                          [/usr/local]
  --exec-prefix=PREFIX    install architecture-dependent files in PREFIX
                          [same as prefix]
  --srcdir=DIR            find the sources in DIR [configure dir or ..]
  --program-prefix=PREFIX prepend PREFIX to installed program names
  --program-suffix=SUFFIX append SUFFIX to installed program names
  --program-transform-name=PROGRAM run sed PROGRAM on installed program names
Host type:
  --build=BUILD           configure for building on BUILD [BUILD=HOST]
  --host=HOST             configure for HOST [guessed]
  --target=TARGET         configure for TARGET [TARGET=HOST]
Features and packages:
  --disable-FEATURE       do not include FEATURE (same as --enable-FEATURE=no)
  --enable-FEATURE[=ARG]  include FEATURE [ARG=yes]
  --with-PACKAGE[=ARG]    use PACKAGE [ARG=yes]
  --without-PACKAGE       do not use PACKAGE (same as --with-PACKAGE=no)
  --x-includes=DIR        X include files are in DIR
  --x-libraries=DIR       X library files are in DIR
--enable and --with options recognized:
  --with-g77              use GNU Fortran 77
  --with-libpath=PATH     use PATH for libraries
  --with-cernlib          use CERNLIB
  --enable-noweb          use noweb(1) to rebuild Fortran sources
  --enable-bison          use bison(1) to rebuild parser
  --enable-paper-a4       use European (A4) paper
  --enable-paper-us       use US (letter) paper
\end{verbatim}}
\caption{\label{fig:configure}%
  Comandline options of the \texttt{configure} script for
  \hepawk/ on UNIX systems.}
\end{figure}

\subsection{Non-UNIX Systems}

For non-UNIX systems configuration and compilation has to be performed
manually:
\begin{itemize}
  \item{} Copy the file \texttt{config.f.in} to \texttt{config.f} and
    edit it according to the comments.
  \item{} If necessary, adapt the \texttt{include} statements in
    \texttt{hepawk.f} and \texttt{hepawk.tab.f} to your \f77/ compiler.
  \item{} Compile the files \texttt{hepawk.f}, \texttt{config.f} and
    \texttt{hepawk.tab.f}.
  \item{} Compile the file \texttt{hktest.f} and link the result with
    the other object files to create a simple test program.
\end{itemize}

\end{document}